\documentclass[9pt,a4paper]{article}
\usepackage{blindtext}
\usepackage[T1]{fontenc}
\usepackage[utf8]{inputenc}
\DeclareUnicodeCharacter{2212}{-}
\usepackage{amsmath}
\usepackage{graphicx}
\graphicspath{{images/}}
\usepackage{subfigure}
\usepackage{hyperref}
\usepackage{authblk}
\usepackage[normalem]{ulem}
\hypersetup{
    colorlinks=true,
    linkcolor=blue,
    filecolor=magenta,      
    urlcolor=cyan,
}
\usepackage{vmargin}
\setmarginsrb{2.5 cm}{1.5 cm}{2.5 cm}{1.5 cm}{0.5 cm}{0.5 cm}{0.5 cm}{0.5 cm}
\date{\today}
\usepackage{setspace}
\usepackage{color,soul}
\title{Hot crystals of thermo-responsive particles with temperature dependent diameter in presence of a temperature gradient}
\author{Rahul Karmakar}
\affil[1]{Department of Physics of Complex Systems, S. N. Bose National Centre for Basic Sciences, Block-JD, Sector-III, Salt Lake Kolkata 700106, India. \\
rahul.physics2017@gmail.com}

\author[1]{J Chakrabarti}

\affil[1]{Department of Physics of Complex Systems, S. N. Bose National Centre for Basic Sciences, Block-JD, Sector-III, Salt Lake Kolkata 700106, India.\\
jaydeb@bose.res.in}%

\begin{document}
\maketitle

\begin{abstract}

Structure formation in non-equilibrium steady state conditions is poorly understood. Non-equilibrium steady state can be achieved in a system by maintaining temperature gradient. A class of cross-linked micro-gel particles, poly-N-isopropylacrylamide (PNIPAM, are reported to  increase in size due to adsorption of water as temperature decreases. Here we study thermo-responsive particles with temperature sensitive diameter in presence of temperature gradient, using Molecular dynamics simulation with Langevin thermostat. We find long-ranged structural order using bond order parameters in both cold and hot region of the system beyond a certain diameter ratio of the cold and hot particles. This is  due to increase in packing and  pressure in both regions.  Our observations might be useful in understanding ordered structures in extreme conditions of non-equilibrium steady state.

\end{abstract}

\newpage

\section{Introduction}
Structure formation in a system often takes place in non-equilibrium steady state conditions\cite{nikoubashman2017self,C7CP02561K,doi:10.1021/nn501407r,di2009colloidal,duhr2005two,lowen2001colloidal}. The examples are plenty:  laning in binary charged colloids in an electric field \cite{dutta2016anomalous,dutta2020length,dutta2018transient}, systems in presence of gradient in thermodynamics variables, like chemical potential \cite{andersen1990steady,roger2018evaporation}, pressure \cite{trau1997assembly} and temperature \cite{hu2011single,moronshing2018room,D1SM01379C}, phase separation in presence of activity\cite{chari2019scalar} and so on.  The mechanisms of structure formation in non-equilibrium steady state remain largely unexplained.

Particle migration under temperature gradient, known as thermophoresis or the Soret effect, is well studied in the literature\cite{duhr2006molecules,perez1994brownian,rubi1998simultaneous,burelbach2018thermophoretic}. There are systems, also called thermo-responsive colloids where, apart from the thermal migration, the particle interactions also change in response to the ambient local temperature \cite{hocine2013thermoresponsive,hu2011single,kim2022crystallization,schmidt2007thermoresponsive}. Many of these systems have technological relevance. However, microscopic studies on thermo-responsive colloids are not well reported to the best of our knowledge. Ligand coated noble metal nano-particle systems are thermo-responsive and have been studied in presence of temperature gradient experimentally \cite{moronshing2018room} and theoretically\cite{D1SM01379C}. These studies show that such colloids form large clusters with long ranged order  which are useful for surface enhanced Raman spectroscopic measurements. Poly-N-iso-propylacrylamide (PNIPAM) micro-gels  are also thermo-responsive colloids. These particles show potential applications in drug delivery\cite{ashraf2016snapshot}, tissue engineering, cell culture and photonic crystal\cite{cui2021aggregated}. The cross-linked  entangled polymer network in PNIPAM traps a large amount of water. The diameter of the  PNIPAM particles increases due to trapped water at low temperature. When heated, they release water resulting in shrinkage of the particles in temperature window between $27^{0}C-34^{0}C$\cite{karthickeyan2017fcc, kolker2021interface}. Experiments show that the size change of the particles can be tuned by co-polymerization of PNIPAM with hydrophilic or hydrophobic species\cite{zhang2006thermosensitive}. PNIPAM particles are experimentally observed to form ordered structures at low temperature due to increase in size\cite{hu2011single} in presence of temperature gradient where the primary focus has been to study crystal-fluid interfacial dynamics. The steady state structural changes of PNIPAM particles in presence of temperature gradient remain unexplored. Since packing lies at the heart of fluid-crystal phase transition \cite{hansen2013theory}, this system pedagogically gives an opportunity to study the interplay between packing and thermal drive to set up structural changes.

Here we model PNIPAM system by a system of particles interacting via the Lennard-Jones (LJ) interaction potential\cite{smit1992phase} with temperature dependent length paratmter $\sigma$. We study the model system using the Langevin dynamics simulations\cite{allen2017computer} in a box with the periodic boundary conditions in three directions. We ignore the hydrodynamic interactions which is a reasonable approximation for small volume fractions. We equilibrate the system at hot temperature and then we cool two ends leaving the middle region of the equilibrium system at the hot temperature, as shown in Fig. \ref{fig:d1 }(a). We study the system for various size ratios of the cold and hot particles at a given  temperature gradient. We characterize the structural order in the system by bond order parameter\cite{wolde1996simulation,doi:10.1021/acs.jctc.6b01034} in the steady state. We observe that beyond a certain size ratio,  not only the  cold region but the hot region also crystallizing due to interplay between packing in different regions and the thermal drift.  

\section{Model and Simulation details}
The interaction potential between a pair of particles is given by: 
\begin{equation}\label{eq:d2}
V^{\alpha}(r)=\epsilon[(\frac{\sigma(T)}{r})^{12}-(\frac{\sigma(T)}{r})^{6}]. 
\end{equation}
Here $\alpha$ (=hot, H and cold, C) denotes the temperature regions in the simulation box. $\epsilon$ is the strength of the potential which is taken to be independent of temperature and $\sigma(T)$ the diameter depending on the local temperature. The particle diameter $\sigma(T)=\sigma_{H}$ for $T=T_{H}$ and $\sigma(T)=\sigma_{C}$ for $T=T_{C}$ and $\sigma_{H}<\sigma_{C}$. $V^{(\alpha)}(r_{ij})$ is total potential energy of the ith particle in the $\alpha$ region with all other particles j at a distance $r_{ij}(=|\vec{r}_{i}-\vec{r}_{j}|)$. For cross interaction between two different sized particles, we use $(\sigma_{H}+\sigma_{C})/2$.

The particle dynamics are computed via the under damped Langevin equation of motion of the ith particle  with mass $m$ at position  $\vec{r_{i}}(t)$ at time t:
\begin{equation}\label{eq:d1}
m_{i} \frac{d^{2} \vec{r}_{i,\alpha}}{dt ^{2}} = -\zeta \frac{d\vec{r}_{i,\alpha}}{dt} - \nabla_{i} \sum_{j=1}^{N} V^{(\alpha)}(r_{ij}) + \vec{f}_{i,\alpha}(t) 
\end{equation}
 $N$ is the total number of particles in the system. $\zeta$ is the friction coefficient. We ignore the temperature dependence of $\zeta$ in our calculation, following our previous work that shows that temperature dependence of viscosity is not important for structure formation in presence of temperature difference\cite{D1SM01379C}.  The components of $\vec{f_{i,\alpha}}(t)$ are the Gaussian white noise with zero mean and variance, $6 \zeta k_{B}T_{H} \delta(t^{'}-t^{''})$ at hot temperature and $6 \zeta k_{B}T_{C} \delta(t^{'}-t^{''})$ in cold temperature, $k_{B}$ being the Boltzmann constant. 

In our simulations, $\epsilon$ is the unit of energy, $\sigma_{H}=0.5 \mu m$  the length unit and $\tau (=\sqrt{\frac{m \sigma_{H}^{2}}{\epsilon}}) \sim 0.2 $ millisecond the time unit for density $1.07 g/cm^{3}$ \cite{saunders2004structure}. We set $\zeta=100$. The discretized equations of motion  are integrated with time step $ 0.001 \tau_{H}$. We perform simulation on $N(=4000)$ colloidal particles in a volume (V) of rectangular parallelepiped box of length $L_{x}=$57.1 and $L_{y}=L_{z}=$10 with the periodic boundary conditions in all three directions at the packing fraction $\eta= \frac{\pi}{6}\frac{N}{V}\sigma_{H}^{3}=$0.36.

\begin{figure}[!htb]
	\centering
	{\includegraphics[width=12.3cm,height=8.3cm]{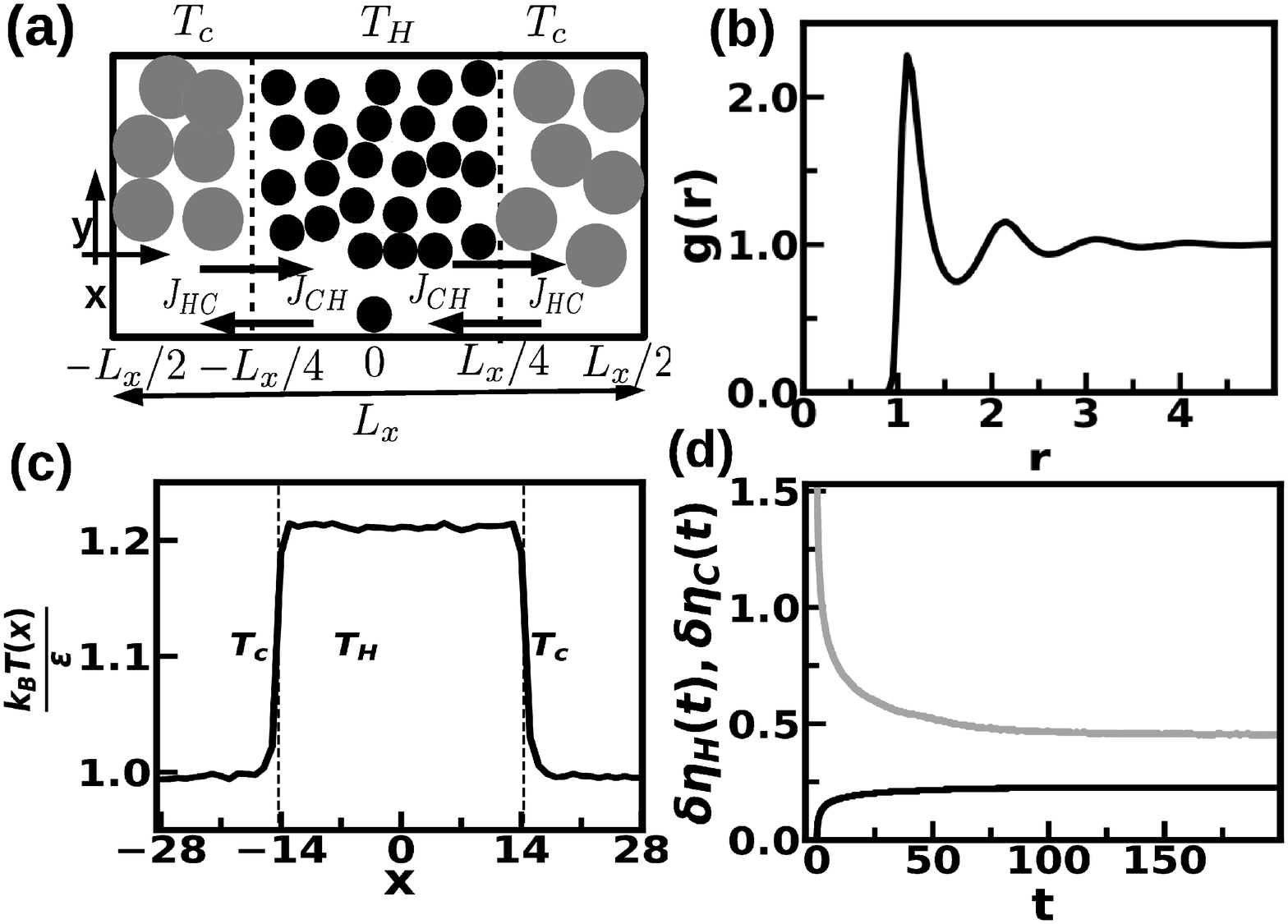}}

	\caption{(a) Schematic diagram of the simulation box after creating temperature gradient: $-L/4<x<L/4$ is the hot region $T_{H}$ and rest of the two side are in cold region $T_{C}$. The small black circles are particles in the hot region. The grey circles show swollen large size particles in the cold region. (b) Pair correleation function $g(r)$ over $r$ in equilibrium at $T_{H}$. (c) $\frac{k_{B}T(x)}{\epsilon}$ versus $x$ plot at steady state after creating temperature gradient for $\sigma^{*}=1.8$. (d) $\delta \eta_{H}$, $\delta \eta_{C}$  versus t for hot region(black) and cold region(grey) for $\sigma^{*}=1.8$.}
	\label{fig:d1 }
\end{figure}

At first for a given $\sigma_{H}$, the system is equilibrated at $T_{H}$.  We create temperature gradient along the x direction. A schematic diagram is given in Fig. \ref{fig:d1 }(a). We thermostat the region (i) $-L_{x}/4$ $<$ x $<$ $L_{x}/4$ with temperature $T_{H}$ and (ii) regions $-L_{x}/2$ $<$ x $<$ $-L_{x}/4$ and $L_{x}/4$ $<$ x $<$ $L_{x}/2$ with temperature $T_{C}$ as depicted in Fig. \ref{fig:d1 }(a). The particle diameter changes according to the temperature of the region as soon as they enter from one region to other. We vary ratio of diameter in cold and hot particles $\sigma^{*}=\frac{\sigma_{C}}{\sigma_{H}}$ for a given $T^{*}=\frac{T_{H}}{T_{C}}$. All the time dependent quantities are averaged over three different trajectories. We take both cold regions for considering data pertaining to the cold region. We calculate steady state quantities averaging over several steady state configurations and also three independent trajectories. Time dependent quantities are calculated by averaging over three independent trajectories. To check the robustness of our observations we also carry out further simulations using different protocols: (1) Constant pressure simulation using steady state pressure from constant volume simulation; (2) the finite size effect; and (3) inclusion of  hydrophobic interaction in the hot region.

\section{Results and Discussions}
We first consider the system in equilibrium at $\frac{k_{B}T_{H}}{\epsilon}=1.2$ and $\eta_{eq}=0.36$. We characterize the equilibrium structure by the radial distribution function $g(r)$. This is the probability  distribution of separations between different pairs of particles over different equilibrium configurations\cite{allen2017computer}. The $g(r)$ data  in Fig. \ref{fig:d1 }(b) shows short-ranged liquid order. 
Next, we create temperature gradient in the equilibrated system as shown schematically in Fig. \ref{fig:d1 }(a) where we take $T^{*}=1.2$ and $\sigma^{*}=1.8$. We plot temperature profile $\frac{k_{B} T(x)}{\epsilon}$(Fig. \ref{fig:d1 }(c) obtained from the average kinetic energy of the particles in a bin width of $\sigma_{H}$ along the x direction in steady state. We calculate the time dependent changes in the packing fraction (Supplementary Material (SM) for details) with respect to the equilibrium condition $\eta_{eq}$, $\delta\eta_{H}(t) = \eta_H (t)-\eta_{eq} $ in hot region and similar data for the cold region $\delta\eta_{C}(t) = \eta_C(t) -\eta_{eq} $. Here $\eta_{H}(t)$ and $\eta_{C}(t)$ are the packing fractions at time $t$ in the hot and the cold region respectively. We show the data in Fig. \ref{fig:d1 }(d). $\delta\eta_{H} (t)$ increases initially and then saturates which we take an indication for setting up the steady state in the system. On the other hand, $\delta\eta_{C}(t)$ decreases for low times and then saturates. The steady values for both cases reach around $25 \tau$.

\subsection{\it Structural Changes}
\begin{figure}[!htb]
	\centering
	{\includegraphics[width=12.3cm,height=8.3cm]{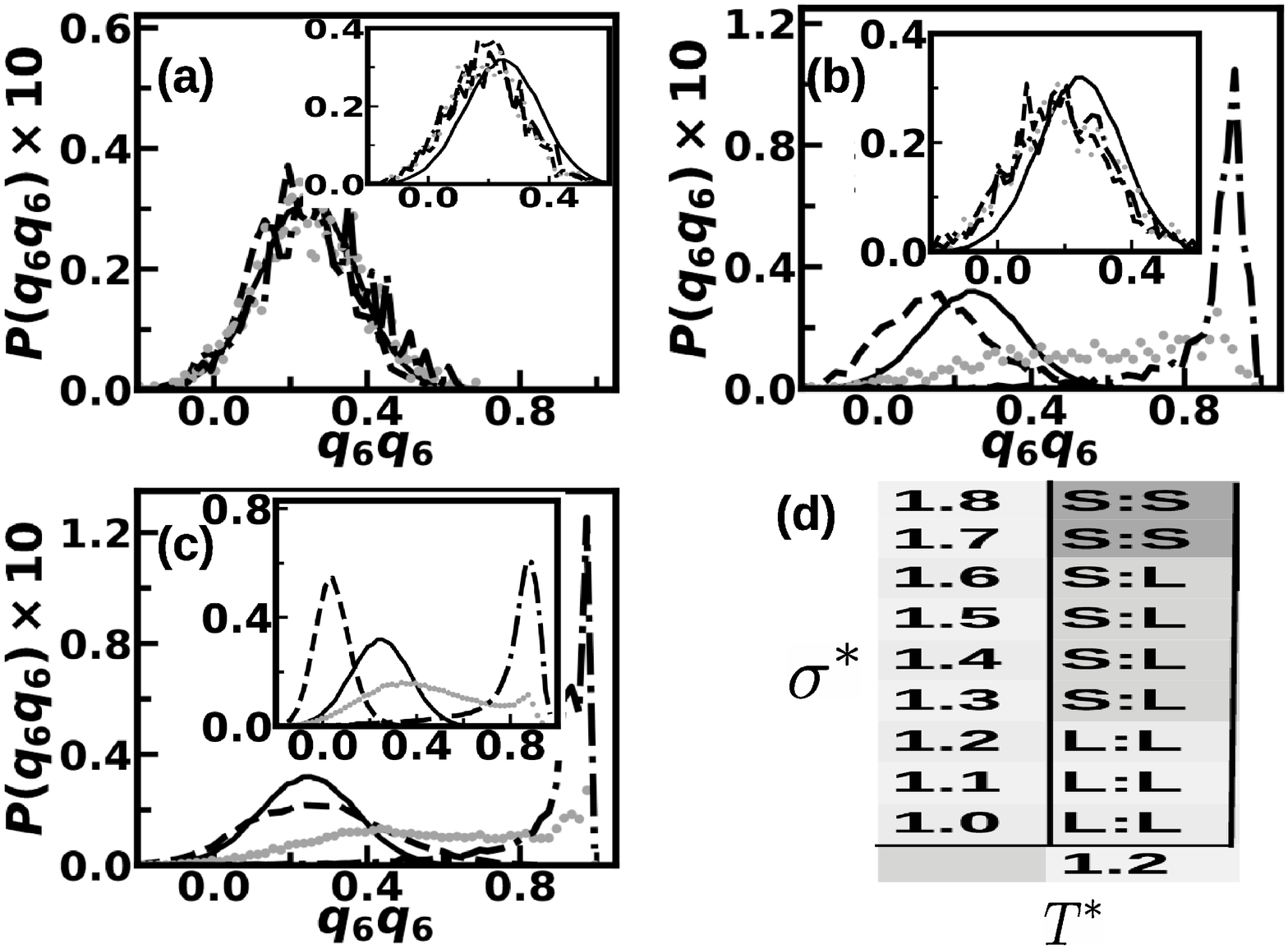}}

	\caption{ (a) $P (q_{6} q_{6} )$ distribution at Equilibrium(solid line), t = $0.5\tau$ (dashed line),$1000\tau$
		(grey line), $4000\tau$ (mixture of black dashed and dotted line) after creating temperature gradient: (a) cold region and inset hot region for $\sigma^*=1.2$. (b) cold region and inset hot region for $\sigma^*=1.6$. (c) cold region and inset hot region for $\sigma^*=1.8$. (d) Steady state structural diagram in $\sigma^{*}-T^{*}$ plane. Symbols L and S denote liquid and ordered state respectively. Left side denotes the structure of cold region, right side for the hot region.}
	\label{fig:d2 }
\end{figure}

 We characterize the temporal evolution of the structure in terms of the distribution of bond orientation order parameter of order $l (=6)$ defined in Ref\cite{doi:10.1021/acs.jctc.6b01034}. To calculate $q_{6}q_{6}$(detailed in SM Eq. 1) we chose neighbour particles with cut off from the minimum after the first peak of $g(r)$ in the equilibrium condition for hot region. For cold region, we chose cut off 1.3 times of the diameter in the cold region. We construct histogram $P(q_{6}q_{6})$ of $ q_{6}q_{6}(i)$ values for the particles in both cold and hot region at different times as shown in Fig. \ref{fig:d2 }(a). We find that in equilibrium at $T_{H}$, $P(q_{6}q_{6})$ has peak around 0.3, which is indicative of a  disordered liquid phase\cite{doi:10.1021/acs.jctc.6b01034}, consistent with the data in Fig. \ref{fig:d1 }(b). 

Let us now consider the case of low $\sigma^*$(=1.2). In the cold region we observe the $P(q_{6}q_{6})$ has a peak around 0.3 for all times indicating liquid order as shown in Fig. \ref{fig:d2 }(a). In inset of Fig. \ref{fig:d2 }(a) we also observe that the hot region remains liquid for all time. The structural  changes occur at $\sigma^*$=1.6. We consider the cold region. Here $P(q_{6}q_{6})$ has peak around 0.3 at low time $t$(= 0.5 $\tau$) in Fig. \ref{fig:d2 }(b). At an intermediate time $ t (=1000 \tau)$, $P(q_{6}q_{6})$ shows broad peak, suggesting coexisting liquid and crystal orders.  At larger time $t (=4000 \tau)$, $P(q_{6}q_{6})$ is sharply peaked at 0.9\cite{doi:10.1021/acs.jctc.6b01034} as for crystal order, with a long tail extended to low $q_{6}q_{6}$. On other hand, the hot region remains liquid for all times as shown in inset of Fig. \ref{fig:d2 }(b). Further structural changes are observed at a higher $\sigma^*$(=1.8) ((Fig. \ref{fig:d2 }(c)). Here
$P(q_{6}q_{6})$ in cold regions behaves as in the earlier case (Fig. \ref{fig:d2 }(b)) indicating crystal order at large times. We observe similar changes of $ P(q_{6}q_{6})$ in the hot region as in the cold region with time in inset of Fig. \ref{fig:d2 }(c). At steady state $P(q_{6}q_{6})$ is sharply peaked around 0.85. Thus both the hot and cold regions show long ranged order. We summarize the steady state structural crossover in Fig. \ref{fig:d2 }(d). In Fig. \ref{fig:d2 }(d), the left letter and the right letter (L=liquid, S=ordered) denotes the structure of cold and hot region respectively. We observe that for  $\sigma^{*}$ up to 1.2, both cold and hot region remains liquid in the steady state. The cold region shows long ranged order along with liquid order in the hot region at $\sigma^{*}=1.3$. Both hot and cold regions show long ranged order after $\sigma^{*}=1.6$.

\begin{figure}[!htb]
	\centering
	{\includegraphics[width=12.3cm,height=8.3cm]{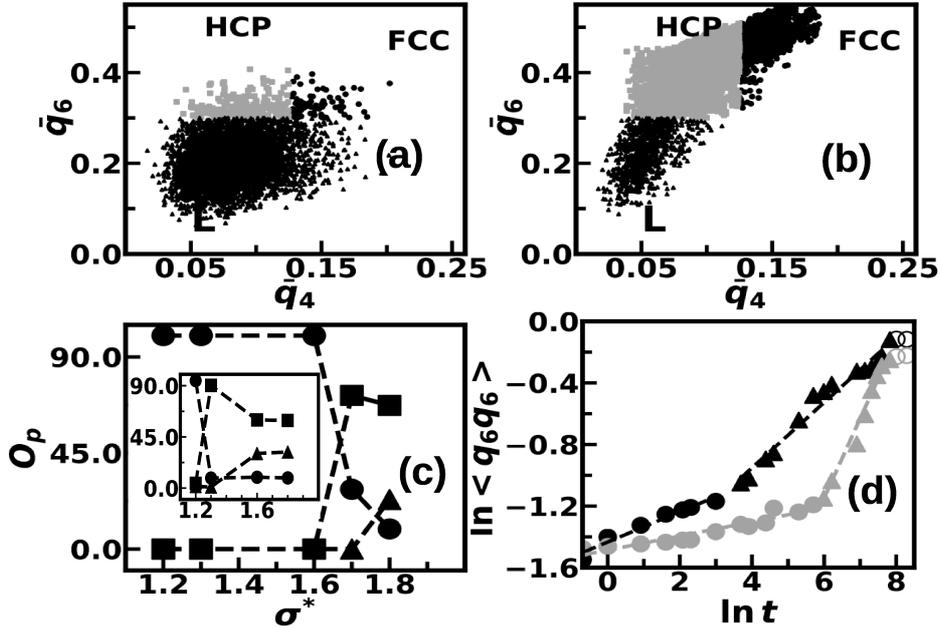}}

	\caption{Scatter plot in $ \bar{q}_{4}- \bar{q}_{6}$ plane at steady state in cold region. The black circles denote FCC structure, grey squares show HCP structure and black triangles are the liquid structure: (a) $\sigma^{*}=1.2$ (b) $\sigma^{*}=1.6$. (c) Percentage $O_{p}$ of different order versus $\sigma^{*}$ in hot region. Liquid (circle), HCP(square), FCC(triangle). The dotted lines are guides to the eyes. Inset: Percentage $O_{p}$ of different order versus $\sigma^{*}$ in cold region. The symbols and axis labels are same as main panel. (d) $\langle q_{6} q_{6} \rangle$ versus time in log-log plot in cold region for $\sigma^{*}=1.8$: black close circle for low time, black triangle for intermediate time, black open circle for large time. Hot region data are shown in grey with the same symbol as cold region. The dotted black lines are fitted lines in cold region and the dotted grey lines are fitted lines in hot region for low and intermediate times.}
	\label{fig:d3 }
\end{figure}

We further calculate the  global rotational invariant quantity \cite{lechner2008accurate} $\bar{q}_{l}$(details in SM, Eq. 2.) and construct the scatter plot in $l=4,6$ plane in both cold and hot region to determine the type of structure. Typical scatter plots are shown in Fig. \ref{fig:d3 }(a) and(b). We show in Fig. \ref{fig:d3 }(c) the percentages of liquid, HCP and FCC orders ($O_{p}$) in hot region in the steady state for different $\sigma^{*}$. We identify the ordered structure following the literature values\cite{lechner2008accurate}. We find in Fig. \ref{fig:d3 }(a) for $\sigma^{*}=1.2$ that cold region is mostly liquid(L) order with very small hexagonal closed pack (HCP) and face centred cube(FCC) crystalline orders. We observe hot region in steady state is similar as in the cold region. For larger $\sigma^{*}(=1.6)$(Fig. \ref{fig:d3 }(b)), however, we observe the cold region has coexisting liquid(L), HCP and FCC crystalline orders in the steady state. On the other hand the hot region remains liquid like with very few hexagonal closed pack (HCP) and face centred cube(FCC) crystalline orders in the steady state. We find for even higher $\sigma^{*}=1.8$ that the cold region has coexisting liquid(L), HCP and FCC crystalline orders in the steady state as in the earlier case in Fig. \ref{fig:d3 }(b).  We find that the hot region at steady state shows similar coexistence phases as the cold region also.

The percentage of different order in Fig. \ref{fig:d3 }(c) shows that in hot region for small $\sigma^{*}(=1.2,1.3)$ predominant order is liquid. The scenario remains same for a slightly higher ratio $\sigma^{*}=1.6$. The scenario changes for higher $\sigma^{*}=1.7$. We observe here that HCP order grows. For even larger $\sigma^{*}(=1.8)$, few FCC order also grows along with HCP order in steady state. In the inset of Fig. \ref{fig:d3 }(c) we show ($O_{p}$) in cold region. We observe for small diameter ratio $\sigma^{*}=1.2$, predominant order is liquid. The scenario changes at $\sigma^{*}=1.3$, the cold region has predominantly HCP order. The percentage of FCC ordering increases and that of the HCP ordering decreases as $\sigma^{*}$ further increases. 

We further show log-log $\langle q_{6}q_{6}(t)\rangle$ over $t$ plot in Fig. \ref{fig:d3 }(d) at  $\sigma^*$=1.8. We observe that $\langle q_{6}q_{6}(t) \rangle $ increases in both regions with time. In both regions, the graph shows two regimes of linear time dependence with different slopes. This indicates algebraic dependence $\langle q_{6}q_{6}(t)\rangle \sim t^{\gamma}$ with different $\gamma$ for two regions. In cold region there is slow increase initially with $\gamma=0.1 $ up to $t=100 \tau$, then increase with $\gamma=0.2$ up to $t=2500 \tau$ and finally to saturation at large time. It is consistent with Fig. \ref{fig:d2 }(c) where we observe peak of the curve shifts at large value of $q_{6}q_{6}$ at large time. Similarly in hot region, the initial growth is slow with exponent $\gamma=0.05$ up to $t=1000 \tau$, then increase with exponent $\gamma=0.5$ up to $t=2500 \tau$ and finally saturates. The growth of the order is steeper in the hot region. This is due to larger space for the particles to move in the hot region.

\begin{figure}[!htb]
	\centering
	{\includegraphics[width=12.3cm,height=8.3cm]{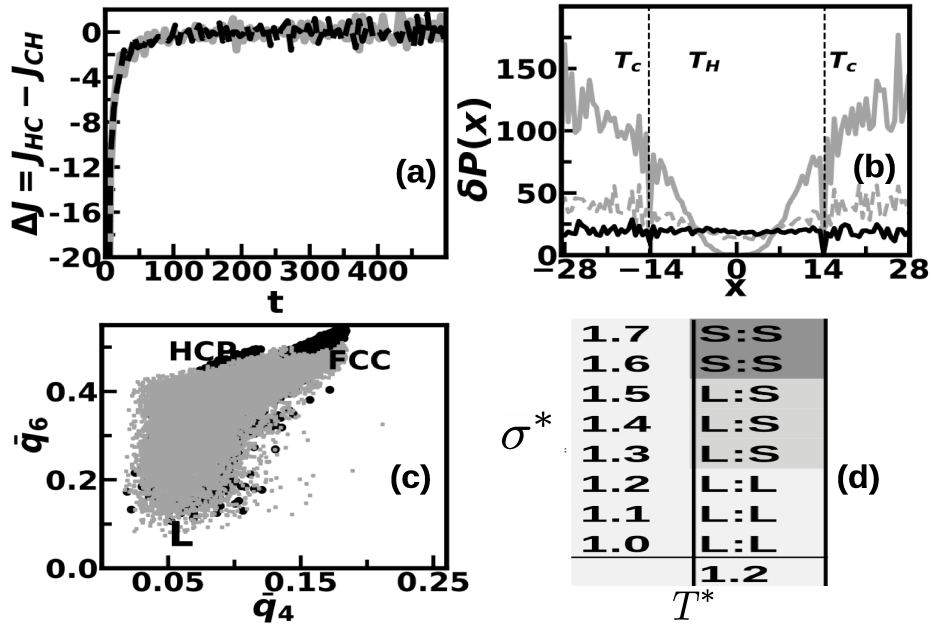}}

	\caption{ (a) Flux $\Delta J = J_{HC}-J_{CH}$ as a function of time in both interfaces for $\sigma^{*}=1.8$. (b) Relative pressure profile $\delta P(x)$ versus $x$ for different time $10\tau$(grey continuous line), $50\tau$(grey dotted line) and steady state(black continuous line). (c) Scatter plot in $ \bar{q}_{4}- \bar{q}_{6}$ plane at steady state in both cold(black circles) and hot(grey circles) region for $\sigma^{*}=1.8$ for larger system size, $N=6400$. (d) Steady state structural diagram in $\sigma^{*}-T^{*}$ plane adding hydrophobic term in the potential. Symbols are the same as described in Fig. \ref{fig:d2 }(d).}
	\label{fig:d4 }
\end{figure}

Physically the emergence of the long ranged order in the hot region in the steady state condition can be understood from the particle flux and the associated change in the packing. We compute the flux $J_{HC}$, given by the number of particles crossing both the interfaces in Fig. \ref{eq:d1}(c) from hot to cold region in a given time interval ($0.2\tau$) and the opposite flux  $J_{CH}$ as well. The flux directions are shown by arrows in Fig. \ref{fig:d1 }(a). The  flux difference, $\Delta J = J_{HC}-J_{CH}$ at both interfaces of the system are plotted in Fig. \ref{fig:d4 }(a). The flux increases from negative values and saturates to zero around $50 \tau$. The negative value indicates that $J_{CH}>J_{HC}$ which means that initially  more particles are pushed out from cold region and accumulate in hot region than those crossing from the hot to the cold region. This is due to increase in particle size in the cold region. The packing driven current surpasses the thermal current from hot to cold region, resulting in increased packing in both regions. 

 We measure pressure in equilibrium at $T_{H}$, the average pressure, $P_{eq}=1.2$. In the steady state, $P_{st}=21.5$ for $\sigma^{*}=1.8$ and $T^{*}=1.2$. (See SM for details). We plot the excess pressure profile $\delta P(x)= P(x)-P_{eq}$ along the direction of temperature gradient $x$(Fig. \ref{fig:d4 }(b)). $\delta P(x)$ is larger in cold region than the hot region at low time. This is so because the packing is larger in the cold region than in the hot region at lower time. Subsequently, $\delta P(x)$ falls  in cold region and simultaneously increases in hot region.  Finally the pressure profile reaches steady state flat value in the whole simulation box with $P_{st}=21.5$. This is also consistent with the time dependence in the packing fractions. The high pressure in both regions in steady state leads to long ranged order.

\subsection{\it{Robustness of the ordered structures}}

 We check if the long-ranged structures both in the cold and hot region is an artifact of the constant volume ensemble that we have simulated.  We perform simulations where we maintain constant pressure(see details in SM) same as the steady state pressure ($P=21.5$) in the system allowing volume of the box to fluctuate, while maintaining the temperature gradient. As volume is fluctuating we scale hot and cold interface region as well to maintain half of the volume hot and other half in cold region. We find that steady state structure in cold and hot region also show ordered state as revealed by the $\bar{q}_{4}-\bar{q}_{6}$ scatter plot (data not shown). Thus, the enhancement of pressure in the steady state helps in experiencing order even in the hot region.

We also study if the finite size effects are important. We simulate a larger system  with box length $L_{x}=75.5, L_{y}=11, L_{z}=11$ and the same particle number density with $N=6400$ particles. Here width of the hot region is $-L_{x}/4$ $<$ x $<$ $L_{x}/4$. Regions $-L_{x}/2$ $<$ x $<$ $-L_{x}/4$ and $L_{x}/4$ $<$ x $<$ $L_{x}/2$ are cold region. We use the parameter $\sigma^{*}=1.8$, $T^{*}=1.2$. Here, we get similar structurally ordered phases in both hot and cold regions shown by the $\bar{q}_{4}-\bar{q}_{6}$ scatter plot in Fig. \ref{fig:d4 }(c). Thus, the long-ranged order formation in the hot and cold region is not sensitive to system size. 

We use another protocol to change the particle diameter in  response to the local temperature  in the system. We estimate that  the velocity auto-correlation function(see details in SM) at $T_{H}$ decay in $0.1 \tau$ ($\sim 100$ simulation steps). We measure temperature profile $T(x)$ from kinetic energy of the particles in a bin width of $\sigma_{H}$ along the x direction, averaging over 100 steps.  We change the diameter linearly with slope $\frac{\sigma_{H}-\sigma_{C}}{T_{H}-T_{C}}$ as per the temperature profile after every 100 simulation steps. We observe both the hot and cold region form long ranged order from $\sigma^{*}=1.5$ (data not shown) for $T^{*}=1.2$, qualitatively similar to earlier results.

It is reported\cite{burmistrova2011effect} that increasing temperature above critical temperature around $32^{0}$ hydrophobic interaction dominates between PNIPAM particles which leads to collapsed state. We include a harmonic potential of spring constant $\beta$ between two particles mimicking hydrophobic attraction as in Ref \cite{chakrabarti2015analytical}, $V_{hyd} = \beta (r-\sigma_{H})^{2}$ with $\beta=0.5$ in addition to the LJ potential in hot region, 
and no such term in the cold region. We equilibrate at hot temperature before creating temperature gradient and observe liquid structure in equilibrium at hot temperature. We repeat our simulation for different $\sigma^{*}$ keeping $T^{*}=1.2$ again. Here we show the steady state structures in Fig. \ref{fig:d4 }(d). We observe both regions remain liquid up to $\sigma^{*}=1.2$ at steady state. At $\sigma^{*}=1.3$, we observe long range order in hot region where as cold region remains liquid, unlike the previous case. The hydrophobic attraction helps in stabilising the ordered phase in the hot region. Upon further increase in $\sigma^{*} (\ge 1.5)$, we observe both regions show long range order at steady state.

Recent experiments\cite{moronshing2018room} on ligand capped metal nano-particles, having temperature sensitive electrostatic potential, show formation of large clusters in the colder region. Ligand coated metal nano-particles have been modelled \cite{D1SM01379C} considering the temperature dependent electrostatic and ligand mediated interactions. The model shows long ranged order in the cold region in steady state in presence of temperature gradient. In this system the long-ranged order is stabilized primarily by slow dynamics due to large packing and slow diffusion in the cold region. This is qualitatively different from the simultaneous crystallization in both hot and cold regions reported here. Many of the non-equilibrium steady state structures are qualitatively different from their equilibrium counterparts\cite{fan20221d}. Setting up of long range order at hot condition is not observed unless pressure is extremely high. Here we encounter crystallization in both regions takes place due to order of magnitude enhanced steady state pressure in the system in presence of competing particle drive and thermal currents. The nonequilibrium steady state condition here supports huge enhancement of pressure. Structural order in extreme pressure conditions are important in other contexts as well. One such example is interior of earth, which is in a constant out-of-equilibrium place where temperature and pressure are very high. The earth's core is abundant of crystalline materials like alumina, silica and iron \cite{lin2004crystal,oganov2005structural,matsui1997case}. 

\section{Conclusions}
The most intriguing result that we report here is the formation of crystalline order even in the hot region in addition to the cold region for sufficiently large $\sigma^{*}$. The hot crystals are stabilized due to an order of magnitude increase in the steady state pressure. Our model illustrates a simple experimentally verifiable system where high temperature crystallization can be studied in the laboratory. This may provide insight to mechanism of formation of long-ranged order in extreme conditions. 

\section*{Supplementary Material}

Supplemental Material contains a brief description of (a) Time dependent packing fraction, (b) Structural Quantities, (c) Pressure calculation in the system, (d) Constant Pressure simulation

\section*{Acknowledgements}
R.K thanks DST Inspire Fellowship (award no. IF170908) for financial support. The authors thank the Thematic Unit of Excellence(TUE) and the Technical Research Centre (TRC) at S.N.Bose National Centre for Basic Sciences for computational facilities.

\bibliographystyle{vancouver}
\bibliography{paper.bib}

\end{document}